\documentclass[12pt]{article}
\usepackage{amsfonts}
\usepackage{latexsym,amsmath,amssymb}
\usepackage{graphics,epstopdf}
\usepackage[pdftex]{graphicx}
\usepackage{subcaption}
\usepackage[export]{adjustbox}

\numberwithin{equation}{section}

\newtheorem{thm}{Theorem}[section]
\newtheorem{cor}[thm]{Corollary}

\numberwithin{equation}{section}

\begin{document}

\bigskip

\bigskip

\begin{center}
{\Large \textbf{B$\acute{e}$zier curves based on Lupa\c{s} $(p,q)$-analogue of
Bernstein polynomials in CAGD
 }}

\bigskip

\textbf{Khalid Khan} and \textbf{D.K. Lobiyal}

School of Computer and System Sciences, SC \& SS, J.N.U., New Delhi-110067., India%
\\[0pt]

 khalidga1517@gmail.com; dklobiyal@gmail.com \\[0pt]

\bigskip

\bigskip

\textbf{Abstract}
\end{center}

\parindent=8mm {\footnotesize {In this paper, we use the blending functions of Lupa\c{s} type (rational) $(p,q)$-Bernstein operators based on $(p,q)$-integers for construction of Lupa\c{s} $(p,q)$-B$\acute{e}$zier curves (rational curves) and surfaces (rational surfaces) with shape parameters. We study  the nature of degree elevation and degree reduction for Lupa\c{s} $(p,q)$-B$\acute{e}$zier Bernstein functions. Parametric curves are represented using Lupa\c{s} $(p,q)$-Bernstein basis.\\

 We introduce affine de Casteljau algorithm for Lupa\c{s} type $(p,q)$-Bernstein B$\acute{e}$zier curves. The new curves have some properties similar to $q$-B$\acute{e}$zier curves. Moreover, we construct the corresponding tensor product surfaces over the rectangular domain $(u, v) \in [0, 1] \times [0, 1] $ depending on four parameters. We also study the de Casteljau algorithm and degree evaluation properties of the surfaces for these generalization over the rectangular domain. Furthermore, some fundamental properties for Lupa\c{s} type $(p,q)$-Bernstein B$\acute{e}$zier curves and surfaces are discussed. We get $q$-B$\acute{e}$zier surfaces for $(u, v) \in [0, 1] \times [0, 1] $  when we set the parameter $p_1=p_2=1.$  In comparison to  $q$-B$\acute{e}$zier curves and surfaces based on Lupa\c{s} $q$-Bernstein polynomials, our generalization gives us more flexibility in controlling the shapes of curves and surfaces.\\

 We also show that the $(p,q)$-analogue of Lupa\c{s} Bernstein operator sequence $L^{n}_{p_n,q_n}(f,x)$ converges uniformly to $f(x)\in C[0,1]$ if and only if $0<q_n<p_n\leq1$ such that $\lim\limits_{n\to\infty} q_n=1, $ $\lim\limits_{n\to\infty} p_n=1$ and $\lim\limits_{n\to\infty}p_n^n=a,$ $\lim\limits_{n\to\infty}q_n^n=b$ with $0<a,b\leq1.$ On the other hand, for any $p>0$ fixed and $p \neq 1,$ the sequence $L^{n}_{p,q}(f,x)$ converges uniformly to $f(x)~ \in C[0,1]$ if and only if $f(x)=ax+b$ for some $a, b \in \mathbb{R}.$

\bigskip


{\footnotesize \emph{MSC: primary 65D17; secondary 41A10, 41A25, 41A36.}: \newline

\bigskip

\section{Introduction and preliminaries}
\parindent=8mm In 1912, S.N. Bernstein \cite{brn} introduced his famous operators $%
B_{n}: $ $C[0,1]\rightarrow C[0,1]$ defined for any $n\in \mathbb{N}$
and for any function $f\in C[0,1]$
\begin{equation}
B_{n}(f;x)=\sum\limits_{k=0}^{n}\left(
\begin{array}{c}
n \\
k%
\end{array}%
\right) x^{k}(1-x)^{n-k}f\biggl{(}\frac{k}{n}\biggl{)},~~x\in \lbrack 0,1].
\end{equation}
 and named it Bernstein polynomials to prove the Weierstrass
theorem \cite{pp}.\\

 In  computer aided geometric design (CAGD), basis of Bernstein polynomials plays a significant role in order to preserve the shape of the curves or surfaces. The classical B$\acute{e}$zier curves \cite{Bezier} constructed with Bernstein basis functions are one of the most important curves in CAGD \cite{thomas}. Apart from this, Bernstein polynomials has sevaral applications in approximation theory \cite{pp}, geometry and computer science due to its fine properties of approximation \cite{hp}.

 In recent years, generalization of the B$\acute{e}$zier curve with shape parameters has received continuous attention.
 Several authors were concerned with the problem of changing the shape of curves and surfaces, while keeping the
control polygon unchanged and thus they generalized the B$\acute{e}$zier curves in \cite{wcq,hp,rababah}. \\

The rapid development of $q$-calculus \cite{vp} has led to the discovery of new generalizations of Bernstein polynomials
involving $q$-integers \cite{lp,ma1,hp,inter} .

\parindent=8mm In 1987, Lupa\c{s} \cite{lp} introduced the first $q$%
-analogue of Bernstein operator as follows

\begin{equation}\label{e9a}
L_{n,q}(f;x)= \sum\limits_{k=0}^{n}~~ \frac{f \bigg(\frac{[k]_{q}}{[n]_{q}}\bigg) ~\left[\begin{array}{c}
n \\
k%
\end{array}%
\right] _{q}  q^{\frac{k(k-1)}{2}}~x^{k}~(1-x)^{n-k}}{\prod\limits_{j=1}^{n}\{(1-x)+q^{j-1} x\}},
\end{equation}
and investigated its approximating and
shape-preserving properties.\\

 In 1996, Phillips \cite{pl} proposed another $q$-variant of the classical Bernstein operator, the so-called Phillips $q$-Bernstein operators  and attracted lots of investigations.

\begin{equation}
B_{n,q}(f;x)=\sum\limits_{k=0}^{n}\left[
\begin{array}{c}
n \\
k%
\end{array}%
\right] _{q}x^{k}\prod\limits_{s=0}^{n-k-1}(1-q^{s}x)~~f\left( \frac{%
[k]_{q}}{[n]_{q}}\right) ,~~x\in \lbrack 0,1]
\end{equation}
where $B_{n,q}: $ $C[0,1]\rightarrow C[0,1]$ defined for any $n\in \mathbb{N}$
and any function $f\in C[0,1].$ \\

  The $q$-variants of Bernstein polynomials
provide one shape parameter for constructing free-form curves and surfaces, Phillips $q$-Bernstein operator was applied well
in this area. In 2003, Oru\c{c} and Phillips \cite{hp} used the basis functions of Phillips $q$-Bernstein operator for construction of
$q$-B$\acute{e}$zier curves and studied the properties of degree reduction and elevation.

\parindent=8mm  Recently, Mursaleen et al. \cite{mka1} applied first the concept of $(p,q)$-calculus in
approximation theory and introduced  $(p,q)$-analogue of Bernstein
operators based on $(p,q)$-integers. They also introduced and studied approximation properties based on $(p,q)$-integers for
$(p,q)$-analogue of Bernstein-Stancu operators, $(p,q)$-analogue of Bernstein-Kantorovich, $(p,q)$-analogue of Bernstein-Shurer operators, $(p,q)$-analogue of Bleimann-Butzer-Hahn operators and $(p,q)$-analogue of Lorentz polynomials on a compact disk in \cite{zmn,mur8,mka3,mka5}.\\

Let us recall certain notations of $(p,q)$-calculus .\\

For any $p>0$ and $q>0,$ the $(p,q)$ integers $[n]_{p,q}$ are defined by

\begin{equation*}
\lbrack n]_{p,q}=p^{n-1}+p^{n-2}q+p^{n-3}q^2+...+pq^{n-2}+q^{n-1}\\
=\left\{
\begin{array}{ll}
\frac{p^{n}-q^{n}}{p-q},~~~~~~~~~~~~~~~~\mbox{when $~~p\neq q \neq 1$  } & \\
&  \\

[n]_q ,~~~~~~~~~~~~~~~~~~~\mbox{when $p=1$  }& \\

n ,~~~~~~~~~~~~~~~~~~~~~\mbox{ when $p=q=1$  }
\end{array}%
\right.
\end{equation*}
~where  $[n]_q $ denotes the $q$-integers and $n=0,1,2,\cdots$.

The formula for $(p,q)$-binomial expansion is as follows:
\begin{equation*}
(ax+by)_{p,q}^{n}:=\sum\limits_{k=0}^{n}p^{\frac{(n-k)(n-k-1)}{2}}q^{\frac{k(k-1)}{2}}
\left[
\begin{array}{c}
n \\
k%
\end{array}%
\right] _{p,q}a^{n-k}b^{k}x^{n-k}y^{k},
\end{equation*}
$$(x+y)_{p,q}^{n}=(x+y)(px+qy)(p^2x+q^2y)\cdots (p^{n-1}x+q^{n-1}y),$$
$$(1-x)_{p,q}^{n}=(1-x)(p-qx)(p^2-q^2x)\cdots (p^{n-1}-q^{n-1}x),$$\\

where  $(p,q)$-binomial coefficients are defined by
$$\left[
\begin{array}{c}
n \\
k%
\end{array}%
\right] _{p,q}=\frac{[n]_{p,q}!}{[k]_{p,q}![n-k]_{p,q}!}.$$

Details on $(p,q)$-calculus can be found in \cite{jag,mka1,mah}.\\

 The $(p,q)$-Bernstein Operators introduced by Mursaleen et al. \cite{mka1} is as follows:
\begin{equation}\label{ee1}
B_{n,p,q}(f;x)=\frac1{p^{\frac{n(n-1)}2}}\sum\limits_{k=0}^{n}\left[
\begin{array}{c}
n \\
k%
\end{array}%
\right] _{p,q}p^{\frac{k(k-1)}2}x^{k}\prod\limits_{s=0}^{n-k-1}(p^{s}-q^{s}x)~~f\left( \frac{%
[k]_{p,q}}{p^{k-n}[n]_{p,q}}\right) ,~~x\in \lbrack 0,1].
\end{equation}

Note when $p=1,$ $(p,q)$-Bernstein Operators given by \ref{ee1} turns out to be $q$-Bernstein Operators.

Also, we have $(p,q)$-analogue of Euler's identity as:
\begin{align*}
(1-x)^{n}_{p,q}&=\prod\limits_{s=0}^{n-1}(p^s-q^{s}x) =(1-x)(p-qx)(p^{2}-q^{2}x)...(p^{n-1}-q^{n-1}x)\\
&=\sum\limits_{k=0}^{n} {(-1)}^{k}p^{\frac{(n-k)(n-k-1)}{2}} q^{\frac{k(k-1)}{2}}\left[
\begin{array}{c}
n \\
k
\end{array}%
\right] _{p,q}x^{k}
\end{align*}

Another needed formulae, which can be easily derived from  Euler's identity for $\mid \frac{q}{p}\mid  <1$ is:

\begin{equation}\label{euler2}
  \sum\limits_{k=0}^{\infty} \frac{q^{\frac{k(k-1)}{2}} x^k}{(p-q)^k {[k] _{p,q}}!}=\prod\limits_{k=0}^{\infty}~ \{ 1+ {\bigg(\frac{q}{p}\bigg)}^{j-1}~x\}
\end{equation}

Again by some simple calculations and using the property of $(p,q)$-integers, we get $(p,q)$-analogue of Pascal's relation as follows:

\begin{equation}\label{e2}
\left[
\begin{array}{c}
n \\
k%
\end{array}%
\right] _{p,q}= q^{n-k}\left[
\begin{array}{c}
n-1 \\
k-1%
\end{array}%
\right] _{p,q}+ p^{k}\left[
\begin{array}{c}
n-1 \\
k%
\end{array}%
\right] _{p,q}
\end{equation}

\begin{equation}\label{e3}
\left[
\begin{array}{c}
n \\
k%
\end{array}%
\right] _{p,q}= p^{n-k}\left[
\begin{array}{c}
n-1 \\
k-1%
\end{array}%
\right] _{p,q}+ q^{k}\left[
\begin{array}{c}
n-1 \\
k%
\end{array}%
\right] _{p,q}
\end{equation}

Motivated by the idea of $(p,q)$-calculus and its importance in the field of approximation theory given by the Mursaleen et al., we construct Lupa\c{s} type (rational) $(p,q)$-B$\acute{e}$zier curves and surfaces based on  $(p,q)$-integers which is further generalization of $q$-B$\acute{e}$zier curves and surfaces \cite{wcq,hp,inter,ph1}.\\

 In next section, We present a new analogue, i.e, Lupa\c{s} type $(p,q)$-analogue of the Bernstein functions. \\

\section{Construction of Lupa\c{s} $(p,q)$-analogue of the Bernstein functions}


 We set

\begin{equation}\label{e9}
b^{k,n}_{p,q}(t)=\frac{\left[\begin{array}{c}
n \\
k%
\end{array}%
\right] _{p,q} p^{\frac{(n-k)(n-k-1)}{2}} q^{\frac{k(k-1)}{2}}~t^{k}~(1-t)^{n-k}}{\prod\limits_{j=1}^{n}\{p^{j-1}(1-t)+q^{j-1} t\}},
\end{equation}
and $b^{0,n}_{p,q}(t), b^{1,n}_{p,q}(t),........,b^{n,n}_{p,q}(t)$ are the $(p,q)$-analogue of the Lupa\c{s} $q$-Bernstein functions \cite{wcq} of degree $n$ on the interval $[0,1]. $ \\


Also for substitution $u=\frac{t}{1-t}$ where $t\in[0,1)$ and $u \in [0, \infty),$

 and using  Euler's identity for $\mid \frac{q}{p}\mid  <1$ is:

\begin{equation}\label{euler2}
  \sum\limits_{k=0}^{\infty} \frac{q^{\frac{k(k-1)}{2}} x^k}{(p-q)^k {[k] _{p,q}}!}=\prod\limits_{k=0}^{\infty}~ \{ 1+ {\bigg(\frac{q}{p}\bigg)}^{j-1}~x\}
\end{equation}

We define

\begin{equation}\label{e99}
b^{k,\infty}_{p,q}(u)= \frac{q^{\frac{k(k-1)}{2}} u^k} {(p-q)^k ~~ {[k] _{p,q}}! ~~  \prod\limits_{j=0}^{\infty}~ \{ 1+ {\bigg(\frac{q}{p}\bigg)}^{j-1} u\}}
\end{equation}

where $b^{k,\infty}_{p,q}(u)$ are the  $(p,q)$-analogue of limit Lupa\c{s} $q$-analogue of the Bernstein functions \cite{sofia} of degree $n$. \\


 \begin{thm}
  The Lupa\c{s} $(p,q)$-analogue of the Bernstein functions  possess the following properties:\\

(1.) Non-negativity: $b^{k,n}_{p,q}(t)\geq 0,~~~$
 $k = 0, 1, . . . , n,~~~~ t \in [0, 1].$\\

(2.) Partition of unity:

\begin{equation*}
\sum\limits_{k=0}^{n}b^{k,n}_{p,q}(t)= 1,~~~~ t \in [0, 1].
\end{equation*}

(3.) End-point property:
\begin{equation*}
b^{k,n}_{p,q}(0)=\left\{
\begin{array}{ll}
1,~~~~~\mbox{if $k=0$ } &  \\
&  \\
0,~~~~~~~~~~\mbox{$ k \neq 0$} &
\end{array}%
\right.
\end{equation*}

\begin{equation*}
b^{k,n}_{p,q}(1)=\left\{
\begin{array}{ll}
1,~~~~~~~~~\mbox{if $k=n$ } &  \\
&  \\
0,~~~~~~~~~~\mbox{$ k \neq n $} &
\end{array}%
\right.
\end{equation*}

(4.) $(p,q)$ inverse symmetry:

\begin{equation*}
b^{n-k,n}_{p,q}(t)=b^{k,n}_{\frac{1}{p},\frac{1}{q}}(1-t)=b^{n-k,n}_{\frac{1}{q},\frac{1}{p}}(t)
\end{equation*} for k = 0, 1, . . . , n.\\

(5.) Reducibility: when $p = 1,$ formula \ref{e9} reduces to the Lupa\c{s} $q$-Bernstein bases.\\

\end{thm}

\textbf{Note:} From  Euler's identity for $\mid \frac{q}{p}\mid  <1$~, we have:

\begin{equation}\label{euler2}
  \sum\limits_{k=0}^{\infty} \frac{q^{\frac{k(k-1)}{2}} x^k}{(p-q)^k {[k] _{p,q}}!}=\prod\limits_{k=0}^{\infty}~ \{ 1+ {\bigg(\frac{q}{p}\bigg)}^{j-1}~x\}
\end{equation}


 \begin{equation*}
\Longrightarrow \sum\limits_{k=0}^{\infty}b^{k,n}_{p,q}(t)= 1,~~~~ t \in [0, 1).
\end{equation*}

\textbf{Proof:}\\ Properties $1, 3$ and $5$ are obvious. Here we only give the proofs of properties $2 $ and $4.$\\

\textbf{Property 2}:

When $t = 1,$ the conclusion is clear; when $t \neq 1,$ we apply the $(p,q)$ analogue of  Newton’s Binomial formula:\\
Consider
(2)\begin{equation*}
\sum\limits_{k=0}^{n}\left[\begin{array}{c}
n \\
k%
\end{array}%
\right] _{p,q} p^{\frac{(n-k)(n-k-1)}{2}} q^{\frac{k(k-1)}{2}}~t^{k}~(1-t)^{n-k}
\end{equation*}

\begin{align*}
  &=\sum\limits_{k=0}^{n}\left[\begin{array}{c}
n \\
k%
\end{array}%
\right] _{p,q} p^{\frac{(n-k)(n-k-1)}{2}} q^{\frac{k(k-1)}{2}}~(1-t)^{n}{\bigg(\frac{t}{1-t}\bigg)}^k\\   &=\bigg(p(1-t)+qt\bigg)~\bigg(p^2(1-t)+q^2t\bigg)~.........\bigg(p^{n-1}(1-t)+q^{n-1}t\bigg) \\
&=\prod\limits_{s=1}^{n}\bigg(p^{s-1}(1-t)+q^{s-1}t\bigg).
\end{align*}

Hence $$ \sum\limits_{k=0}^{n} b^{k,n}_{p,q}(t)=1$$

\textbf{ Property (4)} To prove this result, we need following relations:

$$[n]_{p,q}=[n]_{q,p}~ \text{and}~ \left[
\begin{array}{c}
n \\
k%
\end{array}%
\right] _{p,q}=\left[
\begin{array}{c}
n \\
k%
\end{array}%
\right] _{\frac{1}{q},\frac{1}{p}} \frac{{(pq)}^{\frac{k(2n-1-k)}{2}}}{(pq)^{\frac{k(k-1)}{2}}}.$$

Consider
\begin{align*}
b^{n-k,n}_{p,q}(t)&=\frac{\left[\begin{array}{c}
n \\
n-k%
\end{array}%
\right] _{p,q} p^{\frac{(k)(k-1)}{2}} q^{\frac{(n-k)(n-k-1)}{2}}~t^{n-k}~(1-t)^{k}}{\prod\limits_{j=1}^{n}\{p^{j-1}(1-t)+q^{j-1} t\}}\\
&=\frac{\left[\begin{array}{c}
n \\
n-k%
\end{array}%
\right] _{p,q} p^{\frac{(k)(k-1)}{2}} q^{\frac{(n-k)(n-k-1)}{2}}~t^{n-k}~(1-t)^{k}}{p^{\frac{(n)(n-1)}{2}}~q^{\frac{(n)(n-1)}{2}}\prod\limits_{j=1}^{n}\{\frac{1}{p^{j-1}}t+ \frac{1}{q^{j-1}}(1-t) \}}~~~~~~~~~~~~\\
&=\frac{\left[\begin{array}{c}
n \\
k%
\end{array}%
\right] _{\frac{1}{q},\frac{1}{p}}~ {\frac{1}{p}}^{\frac{(n-k)(n-k-1)}{2}} {\frac{1}{q}}^{\frac{(k)(k-1)}{2}} ~t^{n-k}~(1-t)^{k}}{\prod\limits_{j=1}^{n}\{\frac{1}{p^{j-1}}t+ \frac{1}{q^{j-1}}(1-t) \}}\\
&=b^{k,n}_{\frac{1}{p},\frac{1}{q}}(1-t)\\
&=b^{n-k,n}_{\frac{1}{q},\frac{1}{p}}(t).
\end{align*}

 The Lupa\c{s} $(p,q)$-Bernstein blending functions for $n=3$ are as follows:

 $$ b_{p,q}^{0,3} = \frac{ p^3 (1-t)^3 }  { (p (1-t) + q t)~~ (p^2 (1-t) + q^2 t) }$$
 $$ b_{p,q}^{1,3} = \frac{(p^2+pq+q^2)~ p t (1-t)^2 }  { (p (1-t) + q t)~~ (p^2 (1-t) + q^2 t) }$$
  $$ b_{p,q}^{2,3} = \frac{(p^2+pq+q^2)~ q t^2 (1-t) } {( (p (1-t) + q t)~~ (p^2 (1-t) + q^2 t) )}$$
   $$ b_{p,q}^{3,3} =\frac{  q^3 t^3 }  { (p (1-t) + q t)~~ (p^2 (1-t) + q^2 t) }$$

\begin{figure}[htb!]
\begin{subfigure}{.5\textwidth}
\includegraphics[width=1\linewidth, height=5cm]{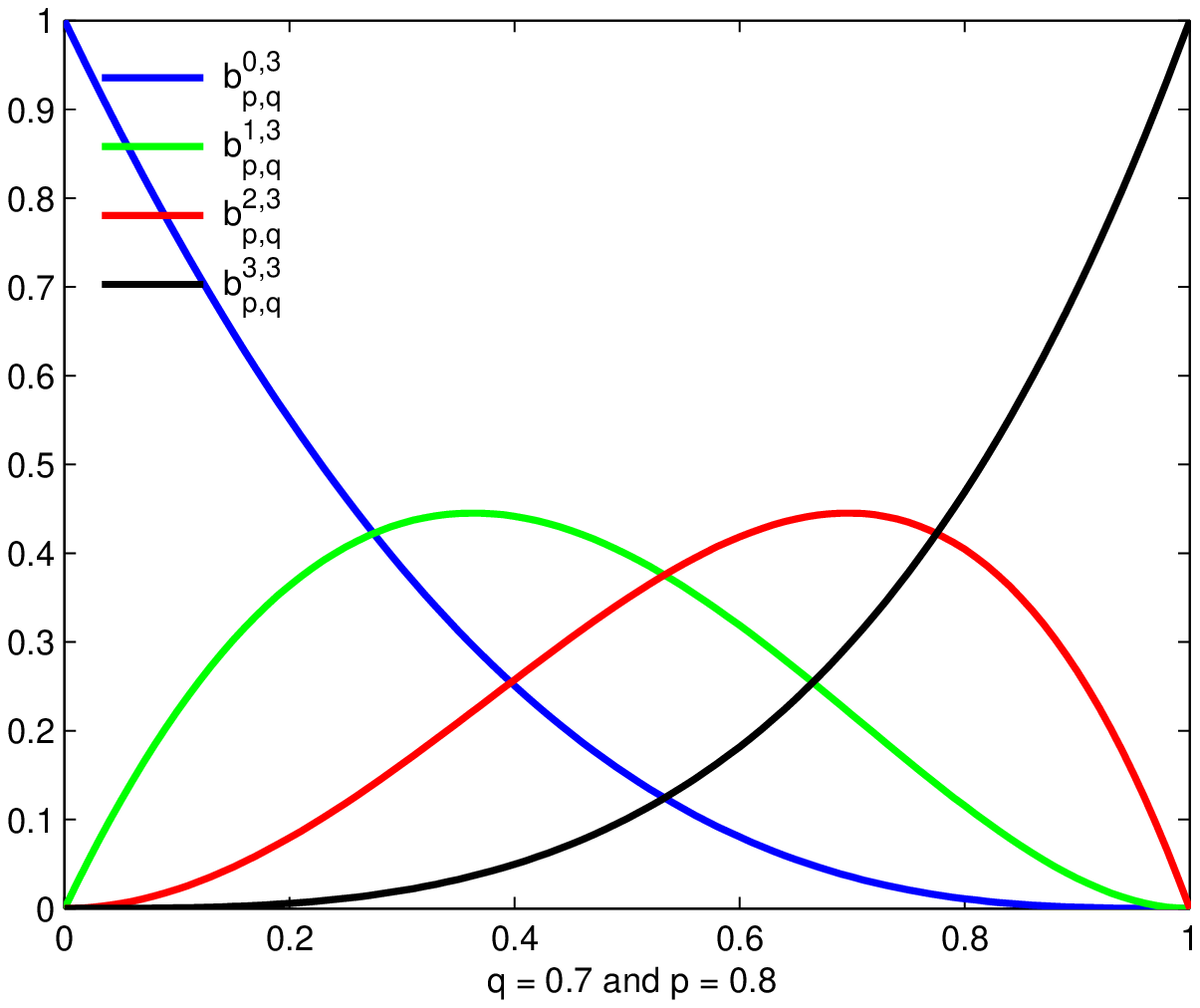}
\caption{$q=0.7, ~p=0.8$ }\label{f3}
\end{subfigure}
\begin{subfigure}{.5\textwidth}
\includegraphics[width=1\linewidth, height=5cm]{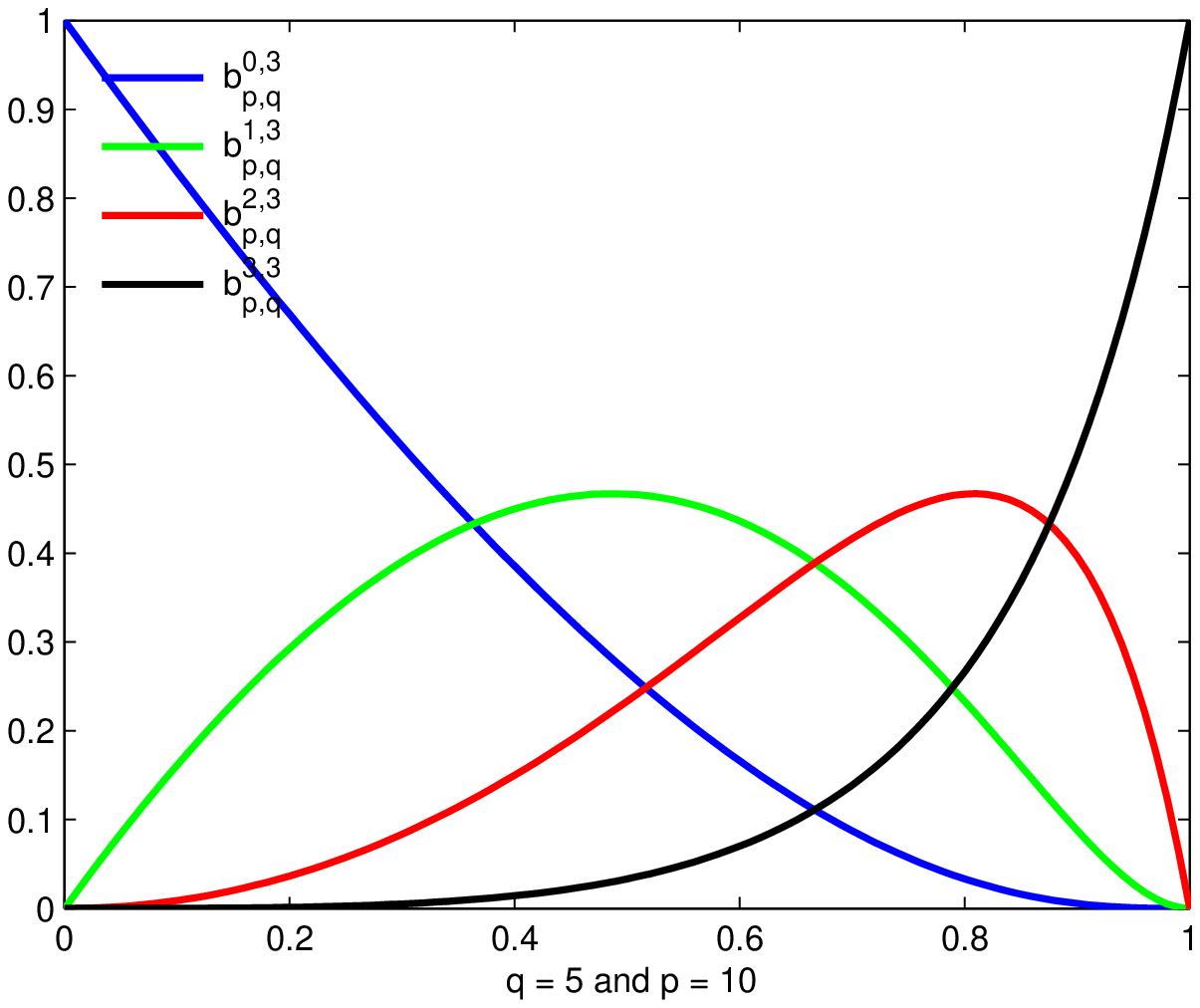}
\caption{$q=5, ~p=10$ }\label{f4}
\end{subfigure}
\caption{Lupa\c{s} cubic B$\acute{e}$zier blending functions}
\label{fig:image5}
\end{figure}

Figure $\ref{f3}$  and $\ref{f4}$ show the Lupa\c{s} $(p,q)$-Bernstein blending functions of degree $3$ for different values of $p$ and $q$ . Here we can observe that sum of blending  fuctions is always unity.

%
%
%

\newpage

\section {Degree elevation and reduction for Lupa\c{s} $(p,q)$-Bernstein functions}

Technique of degree elevation has been used to increase the flexibility of a given curve.
A degree elevation algorithm calculates a new set of control points by choosing a convex combination
of the old set of control points which retains the old end points. For this purpose, the identities (\ref{ee6}),(\ref{ee7}) and Theorem (\ref{lt1}) are useful.\\

\textbf{Degree elevation}

\begin{equation}\label{ee6}
 \frac{q^{n}~t}{p^{n}(1-t)+q^{n}~t}  b^{k,n}_{p,q}(t)=\bigg(1-\frac{p^{k+1}[n-k]}{n+1}\bigg) b^{k+1,n+1}_{p,q}(t)
\end{equation}

\begin{equation}\label{ee7}
 \frac{p^n(1-t)}{p^{n}(1-t)+q^{n}~t}  b^{k,n}_{p,q}(t)=\bigg(\frac{p^{k}[n+1-k]}{n+1}\bigg) b^{k,n+1}_{p,q}(t)
\end{equation}

\begin{thm}\label{lt1}
 Each Lupa\c{s} $(p,q)$-analogue of the corresponding Bernstein function of degree $n$ is a linear combination of two Lupa\c{s}
$(p,q)$-analogues of the Bernstein functions of degree $n + 1:$\\

\begin{equation}\label{e10}
b^{k,n}_{p,q}(t)= \bigg(\frac{p^{k}~{[n+1-k]}_{p,q}}{{[n+1]}_{p,q}}\bigg)~b^{k,n+1}_{p,q}(t)+\bigg(1-\frac{p^{k+1}~{[n-k]}_{p,q}}{{[n+1]}_{p,q}}\bigg)~b^{k+1,n+1}_{p,q}(t)
\end{equation}\\
\end{thm}
\textbf{Proof:}
\begin{align*}
  b^{k,n}_{p,q}(t)&= b^{k,n}_{p,q}(t)\bigg( 1-\frac{q^n~t}{p^n(1-t)+q^n~t}+\frac{q^n~t}{p^n(1-t)+q^n~t}\bigg)  \\
  &= \frac{p^n(1-t)}{p^n(1-t)+q^n~t}\bigg(\frac{\left[\begin{array}{c}
n \\
k%
\end{array}%
\right] _{p,q} p^{\frac{(n-k)(n-k-1)}{2}} q^{\frac{k(k-1)}{2}}~t^{k}~(1-t)^{n-k}}{\prod\limits_{j=1}^{n}\{p^{j-1}(1-t)+q^{j-1} t\}} \bigg)\\
&+\frac{q^n~t}{p^n(1-t)+q^n~t}\bigg(\frac{\left[\begin{array}{c}
n \\
k%
\end{array}%
\right] _{p,q} p^{\frac{(n-k)(n-k-1)}{2}} q^{\frac{k(k-1)}{2}}~t^{k}~(1-t)^{n-k}}{\prod\limits_{j=1}^{n}\{p^{j-1}(1-t)+q^{j-1} t\}} \bigg).
\end{align*}

Using \ref{ee6} and \ref{ee7}, we have

\begin{equation*}
b^{k,n}_{p,q}(t)= \bigg(\frac{p^{k}~{[n+1-k]}_{p,q}}{{[n+1]}_{p,q}}\bigg)~b^{k,n+1}_{p,q}(t)+\bigg(1-\frac{p^{k+1}~{[n-k]}_{p,q}}{{[n+1]}_{p,q}}\bigg)~b^{k+1,n+1}_{p,q}(t)
\end{equation*}\\

\begin{thm}\label{khalidthm}
 Each Lupa\c{s} $(p,q)$-analogue of the Bernstein function of degree $n$ is a linear combination of two Lupaş $(p,q)$-analogues of
the Bernstein functions of degree $n-1:$\\

\begin{equation}\label{e11}
b^{k,n}_{p,q}(t)=\frac{q^{n-1}~t}{p^{n-1}(1-t)+q^{n-1} t}~b^{k-1,n-1}_{p,q}(t)+\frac{p^{n-1}(1-t)}{p^{n-1}(1-t)+q^{n-1} t}~b^{k,n-1}_{p,q}(t)
\end{equation}

\begin{equation}\label{e11a}
b^{k,n}_{p,q}(t)=\frac{p^{n-k}q^{k-1}t}{p^{n-1}(1-t)+q^{n-1} t}~b^{k-1,n-1}_{p,q}(t)+\frac{p^{n-k-1}q^{k}(1-t)}{p^{n-1}(1-t)+q^{n-1} t}~b^{k,n-1}_{p,q}(t)
\end{equation}
\end{thm}

\textbf{Proof} We use the Pascal's type relations of the $(p,q)$-Binomial coefficient.\\
According to formula \ref{e3},

\begin{align*}
b^{k,n}_{p,q}(t)&=\frac{\bigg(p^{n-k}\left[\begin{array}{c}
n-1 \\
k-1%
\end{array}%
\right] _{p,q}+q^k ~\left[\begin{array}{c}
n-1 \\
k%
\end{array}%
\right] _{p,q}\bigg)~ p^{\frac{(n-k)(n-k-1)}{2}} q^{\frac{k(k-1)}{2}}~t^{k}~(1-t)^{n-k}}{\prod\limits_{j=1}^{n}\{p^{j-1}(1-t)+q^{j-1} t\}}\\
\end{align*}
or
\begin{align*}
b^{k,n}_{p,q}(t)&=\frac{p^{n-k}q^{k-1}t}{p^{n-1}(1-t)+q^{n-1}~ t} \frac{\left[\begin{array}{c}
n-1 \\
k-1%
\end{array}%
\right] _{p,q}~p^{\frac{(n-k)(n-k-1)}{2}} q^{\frac{(k-1)(k-2)}{2}}~t^{k-1}~(1-t)^{n-k}}{\prod\limits_{j=1}^{n-1}\{p^{j-1}(1-t)+q^{j-1} t\}}\\  &+\frac{p^{n-k-1}q^{k}(1-t)}{p^{n-1}(1-t)+q^{n-1} t}~\frac{\left[\begin{array}{c}
n-1 \\
k%
\end{array}%
\right] _{p,q}~p^{\frac{(n-1-k)(n-k-2)}{2}} q^{\frac{k(k-1)}{2}}~t^{k}~(1-t)^{n-k-1}}{\prod\limits_{j=1}^{n-1}\{p^{j-1}(1-t)+q^{j-1} t\}}\\
&=\frac{p^{n-1}q^{k-1}t}{p^{n-1}(1-t)+q^{n-1} t}~b^{k-1,n-1}_{p,q}(t)+\frac{p^{n-k-1}q^{k}(1-t)}{p^{n-1}(1-t)+q^{n-1} t}~b^{k,n-1}_{p,q}(t)
\end{align*}

or
\begin{align*}
  b^{k,n}_{p,q}(t)&=\frac{\bigg(q^{n-k}\left[\begin{array}{c}
n-1 \\
k-1%
\end{array}%
\right] _{p,q}+p^k ~\left[\begin{array}{c}
n-1 \\
k%
\end{array}%
\right] _{p,q}\bigg)~ p^{\frac{(n-k)(n-k-1)}{2}} q^{\frac{k(k-1)}{2}}~t^{k}~(1-t)^{n-k}}{\prod\limits_{j=1}^{n}\{p^{j-1}(1-t)+q^{j-1} t\}}\\
&=\frac{q^{n-1}~t}{p^{n-1}(1-t)+q^{n-1} t}~b^{k-1,n-1}_{p,q}(t)+\frac{p^{n-1}(1-t)}{p^{n-1}(1-t)+q^{n-1} t}~b^{k,n-1}_{p,q}(t)
\end{align*}

\section{Lupa\c{s} $(p,q)$-B$\acute{e}$zier curves:}

Let us define the Lupa\c{s} $(p,q)$-B$\acute{e}$zier curves of degree n using the Lupa\c{s} $(p,q)$-analogues of the Bernstein functions as follows:

\begin{equation}\label{e12}
{\bf{ P}}(t; p,q) = \sum\limits_{i=0}^{n} {\bf{P_i}}~ b^{k,n}_{p,q}(t)
\end{equation}

where $P_i \in R^3$  $(i = 0, 1, . . . , n)$ and $ p>q > 0.$ $P_i$ are control points. Joining up adjacent points $P_i,$  $i = 0, 1, 2, . . . , n $ to obtain a
polygon which is called the control polygon of Lupa\c{s} $(p,q)$-B$\acute{e}$zier curves.\\

\subsection{ Some basic properties of Lupa\c{s} $(p,q)$-B$\acute{e}$zier curves.}

 \begin{thm} From the definition, we can derive some basic properties of Lupa\c{s} $(p,q)$-B$\acute{e}$zier curves:\\

\noindent 1. Lupa\c{s} $(p,q)$-B$\acute{e}$zier curves have geometric and affine invariance.\\
2. Lupa\c{s} $(p,q)$-B$\acute{e}$zier curves lie inside the convex hull of its control polygon.\\
3. The end-point interpolation property: ${\bf{P}}(0; p, q) = {\bf{P_0,}} ~ {\bf{P}}(1; p, q) = \bf{P_n.}$\\
4. $(p,q)$-inverse symmetry: the Lupa\c{s} $(p,q)$-B$\acute{e}$zier curves obtained by reversing the order of the control points is the same as the Lupa\c{s}
  $(p,q)$-B$\acute{e}$zier curves with $q$ replaced by $\frac{1}{q} $ and $p$ replaced by $\frac{1}{p} $.\\
5. Reducibility: when $p = 1,$ formula \ref{e12} gives the $q$-B$\acute{e}$zier curves.
\end{thm}

\textbf{Proof.} These properties of Lupa\c{s} $(p,q)$-B$\acute{e}$zier curves can be easily deduced from corresponding properties of the Lupa\c{s}
$(p,q)$-analogue of the Bernstein functions. Here we only give the proof of property 4.\\

Let ${\bf{P_i^\ast}}={\bf{P_{n-i}}},~~$
$ i = 0, 1, . . . , n,$ then we have

\begin{align*}
  {\bf{ P^\ast}}(t; p,q) &= \sum\limits_{k=0}^{n} {\bf{P_i^\ast}}~ b^{k,n}_{p,q}(t) \\
 &= \sum\limits_{k=0}^{n} {\bf{P_i^\ast}}~ b^{k,n}_{\frac{1}{p},\frac{1}{q}}(1-t) \\
 &={\bf{ P}}(1-t; \frac{1}{p},\frac{1}{q}).
\end{align*}

\begin{thm} The end-point property of derivative:
$${\bf{P^{\prime}}}(0; p,q)= \frac{[n]_{p,q}}{p^{n-1}}({\bf{P_1}}-{\bf{P_0}})$$

  $${\bf{P^{\prime}}}(1; p,q)= \frac{[n]_{p,q}}{q^{n-1}}({\bf{P_n}}-{\bf{P_{n-1}}})$$

i.e. Lupa\c{s} $(p,q)$-B$\acute{e}$zier curves are tangent to fore-and-aft edges of its control polygon at end points.
\end{thm}
\textbf{Proof:}
Let

\begin{align}\label{e13}
  {\bf{ P}}(t; p,q) = \sum\limits_{k=0}^{n} {\bf{P_k}}~ b^{k,n}_{p,q}(t)&= \frac{\sum\limits_{k=0}^{n}~{\bf{P_k}}~\left[\begin{array}{c}
n \\
k%
\end{array}%
\right] _{p,q} p^{\frac{(n-k)(n-k-1)}{2}} q^{\frac{k(k-1)}{2}}~t^{k}~(1-t)^{n-k}}{\prod\limits_{j=1}^{n}\{p^{j-1}(1-t)+q^{j-1} t\}}\\
&=\frac{{\bf{V}}(t; p,q)}{{\bf{ W}}(t; p,q)}
\end{align}
 or
 $${\bf{P}}(t; p,q)~ {{\bf{ W}}(t; p,q)}= {\bf{V}}(t; p,q)$$\\
 then on differentiating both hand side with respect to `t', we have\\
 $${\bf{ P^\prime}}(t; p,q)~ {{\bf{ W}}(t; p,q)}+{\bf{ P}}(t; p,q)~ {{\bf{ W^\prime}}(t; p,q)}= {\bf{V^\prime}}(t; p,q).$$\\
 Let $$A_{k}^{n}(t; p,q)~=\left[\begin{array}{c}
n \\
k%
\end{array}%
\right] _{p,q} p^{\frac{(n-k)(n-k-1)}{2}} q^{\frac{k(k-1)}{2}}~t^{k}~(1-t)^{n-k},$$ then

$${\bf{V}}(t; p,q)=\sum\limits_{k=0}^{n}~{\bf{P_k}} A_{k}^{n}(t; p,q)$$\\
From property $2$ of the Lupa\c{s} $(p,q)$-Bernstein functions, we obtain\\
$${\bf{W}}(t; p,q)=\sum\limits_{k=0}^{n}~ A_{k}^{n}(t; p,q)$$\\
as

\begin{align*}
  {(A_{k}^{n}(t; p,q)}^{\prime}&=\frac{[n]_{p,q}}{[k]_{p,q}}\left[\begin{array}{c}
n-1 \\
k-1%
\end{array}%
\right] _{p,q} p^{\frac{(n-k)(n-k-1)}{2}} q^{\frac{k(k-1)}{2}}~k~t^{k-1}~(1-t)^{n-k}\\
 &-\frac{[n]_{p,q}}{[n-k]_{p,q}}\left[\begin{array}{c}
n-1 \\
k%
\end{array}%
\right] _{p,q} p^{\frac{(n-k)(n-k-1)}{2}} q^{\frac{k(k-1)}{2}}~(n-k)~t^{k}~(1-t)^{n-k-1}\\
&=\frac{[n]_{p,q}}{[k]_{p,q}}~q^{k-1}k~{A_{k-1}^{n-1}(t; p,q)}-\frac{[n]_{p,q}}{[n-k]_{p,q}}~p^{n-k-1}(n-k)~{A_{k}^{n-1}(t; p,q)}\\
&=C_{k}^{n}~{A_{k-1}^{n-1}(t; p,q)}- D_{n-k}^{n}~{A_{k}^{n-1}(t; p,q)}
\end{align*}

where
$$C_{k}^{n}= \frac{[n]_{p,q}}{[k]_{p,q}}~q^{k-1}k,~~D_{n-k}^{n}~=\frac{[n]_{p,q}}{[n-k]_{p,q}}~p^{n-k-1}(n-k).$$ \\
Then $${\bf{V}}(0; p,q)= {\bf{P_0}~p^{\frac{n(n-1)}{2}}},~~~~{\bf{W}}(0; p,q)= ~p^{\frac{n(n-1)}{2}}$$

$${\bf{V^{\prime}}}(0; p,q)=( C_{1}^{n}~{\bf{P_1}}-D_{n}^{n}~{\bf{P_0}})~p^{\frac{(n-1)(n-2)}{2}},$$

$${\bf{W^{\prime}}}(0; p,q)= (C_{1}^{n}~-D_{n}^{n}~) ~p^{\frac{(n-1)(n-2)}{2}},$$ hence\\
$${\bf{P^{\prime}}}(0; p,q)= \frac{[n]_{p,q}}{p^{n-1}}({\bf{P_1}}-{\bf{P_0}})$$

Similarly, we have

 $${\bf{V}}(1; p,q)= {\bf{P_n}~q^{\frac{n(n-1)}{2}}},~~~~{\bf{W}}(1; p,q)= ~q^{\frac{n(n-1)}{2}}$$

$${\bf{V^{\prime}}}(1; p,q)=( C_{n}^{n}~{\bf{P_n}}-D_{1}^{n}~{\bf{P_{n-1}}})~q^{\frac{(n-1)(n-2)}{2}},$$

$${\bf{W^{\prime}}}(1; p,q)= (C_{n}^{n}~-D_{1}^{n}~) ~q^{\frac{(n-1)(n-2)}{2}},$$ hence\\
$${\bf{P^{\prime}}}(1; p,q)= \frac{[n]_{p,q}}{q^{n-1}}({\bf{P_n}}-{\bf{P_{n-1}}})$$\\

\begin{thm}\label{tp1}
 Planar Lupa\c{s} $(p,q)$-B$\acute{e}$zier curves are variation diminishing, which means that the number of times any straight line
crosses the Lupa\c{s} $(p,q)$-B$\acute{e}$zier curve is no more than the number of times it crosses the control polygon.\\
\end{thm}

\textbf{Proof.} For any polynomial $f(t),$ we denote $Z_{t\in I\subseteq(0,\infty)}[f (t)]$ as the number of positive roots of $f (t)$ on the interval I. For
any vector $V = (v_0, v_1, . . . , v_n),$ we write $S^{-}(v_0, v_1, . . . , v_n)$ to denote the number of strict sign changes in the vector $V.$

Because the sequence of functions $(1, t, . . . , t^n)$ is totally positive on $[0, 1],$ then for any sequence of real numbers
$a_0, a_1, . . . , a_n,$
$Z_{0 <t <1} [a_0 + a_1 t + · · · + a_n~ t^n] = S^{-}(a_0 + a_1t + · · · + a_n~t^n) \leq S^{-}(a_0, a_1, . . . , a_n).$\\

Let C denote a planar Lupa\c{s} $(p,q)$-B$\acute{e}$zier curve, L any straight line, and let I(C, L) the number of times C crosses L. Establish
the rectangular coordinate system whose abscissa axis is L. Because Lupa\c{s} $(p,q)$-B$\acute{e}$zier curves are geometric invariant, we can
denote $(x_i, y_i) (i = 0, 1, . . . , n)$ the new coordinates of the control points. Let P denote the control polygon and $I(P, L)$ the
number of times P crosses L. Then we will prove that $I(C, L) \leq  I(P, L).$\\

We make a parameter transformation. Let $ u = \frac{t}{1-t},~~~~~t \in (0, 1),$ so that $ u \in (0,\infty).$ Then
\begin{align*}
  {\bf{ I}}(C,L)&= Z_{0<t<1}\bigg[\sum\limits_{k=0}^{n} {\bf{y_k}}~ b^{k,n}_{p,q}(t) \bigg]\\
  &= Z_{0<t<1} \bigg[\frac{\sum\limits_{k=0}^{n}~{\bf{y_k}}~\left[\begin{array}{c}
n \\
k%
\end{array}%
\right] _{p,q} p^{\frac{(n-k)(n-k-1)}{2}} q^{\frac{k(k-1)}{2}}~t^{k}~(1-t)^{n-k}}{\prod\limits_{j=1}^{n}\{p^{j-1}(1-t)+q^{j-1} t\}}
\bigg]\\
&=Z_{0<u<\infty} \bigg[\frac{\sum\limits_{k=0}^{n}~{\bf{y_k}}~\left[\begin{array}{c}
n \\
k%
\end{array}%
\right] _{p,q} p^{\frac{(n-k)(n-k-1)}{2}} q^{\frac{k(k-1)}{2}}~u^{k}~}{\prod\limits_{j=1}^{n}\{p^{j-1}+q^{j-1} u\}}\bigg]\\
&=Z_{0<u<\infty} \bigg[\sum\limits_{k=0}^{n}~{\bf{y_k}}~\left[\begin{array}{c}
n \\
k%
\end{array}%
\right] _{p,q} ~u^{k}~\bigg]\\
&\leq S^{-}\bigg(\left[\begin{array}{c}
n \\
0%
\end{array}%
\right] _{p,q}y_{0},\left[\begin{array}{c}
n \\
1%
\end{array}%
\right] _{p,q}y_{1},.......,\left[\begin{array}{c}
n \\
n%
\end{array}%
\right] _{p,q}y_{n}\bigg)\\
&=S^{-}(y_{0},y_{1}.......,y_{n})
\end{align*}

From \ref{tp1}, we can obtain the following two corollaries:\\

\begin{cor}
Convexity-preserving: the planar Lupa\c{s} $(p,q)$-B$\acute{e}$zier curve is convex, as long as its control polygon is convex.\\
\end{cor}

\begin{cor} Monotonicity-preserving: let the control polygon be monotonically increasing (decreasing) in a given direction,
then the planar Lupa\c{s} $(p,q)$-B$\acute{e}$zier curve is also monotonically increasing (decreasing).
\end{cor}

 \subsection{Degree elevation for Lupa\c{s} $(p,q)$-B$\acute{e}$zier curves}

Lupa\c{s} $(p,q)$-B$\acute{e}$zier curves have a degree elevation algorithm that is similar to that possessed by the classical B$\acute{e}$zier curves.
Using the technique of degree elevation, we can increase the flexibility of a given curve.

 $$ {\bf{ P}}(t; p,q) = \sum\limits_{k=0}^{n} {\bf{P_k}}~ b^{k,n}_{p,q}(t)$$

 $$ {\bf{ P}}(t; p,q) = \sum\limits_{k=0}^{n+1} {\bf{P_k^\ast}}~ b^{k,{n+1}}_{p,q}(t),$$ where

\begin{equation}\label{ee15}
{\bf{ P_k^\ast}}=\bigg(1-\frac{p^{k}~{[n+1-k]}_{p,q}}{{[n+1]}_{p,q}}\bigg)~{\bf{ P_{k-1}}}+ \bigg(\frac{p^{k}~{[n+1-k]}_{p,q}}{{[n+1]}_{p,q}}\bigg)~{\bf{ P_k}}
\end{equation}

The statement above can be derived using the identities $(\ref{ee6}) \text{and} (\ref{ee7}).$
Consider

\begin{equation*}
      {\bf{ P}}(t; p,q)= \frac{p^n(1-t)}{p^{n}(1-t)+q^{n}~t} {\bf{ P}}(t; p,q)+\frac{ q^n ~t}{p^{n}(1-t)+q^{n}~t} {\bf{ P}}(t; p,q)
\end{equation*}

We obtain

\begin{equation*}
{\bf{P}}(t; p,q)=\sum\limits_{k=0}^{n} \bigg(p^k ~~\frac{~{[n+1-k]}_{p,q}}{{[n+1]}_{p,q}}\bigg) {\bf{P^0_k}} b^{k,n+1}_{p,q}(t) + \sum\limits_{k=0}^{n}\bigg(1-\frac{p^{k+1}~{[n-k]}_{p,q}}{{[n+1]}_{p,q}}\bigg){\bf{P^0_k}} b^{k+1,n+1}_{p,q}(t)
\end{equation*}

Now by shifting the limits, we have

\begin{equation*}
{\bf{P}}(t; p,q)=\sum\limits_{k=0}^{n+1} \bigg(p^k ~~\frac{~{[n+1-k]}_{p,q}}{{[n+1]}_{p,q}}\bigg) {\bf{P^0_k}} b^{k,n+1}_{p,q}(t) + \sum\limits_{k=0}^{n+1}\bigg(1-\frac{p^{k}~{[n+1-k]}_{p,q}}{{[n+1]}_{p,q}}\bigg){\bf{P^0_{k-1}}} b^{k,n+1}_{p,q}(t)
\end{equation*}

where ${\bf{P^0_{-1}}} $is defined as the zero vector.
Comparing coefficients on both side, we have

\begin{equation*}
{\bf{ P_k^\ast}}=\bigg(1-\frac{p^{k}~{[n+1-k]}_{p,q}}{{[n+1]}_{p,q}}\bigg)~{\bf{ P_{k-1}}}+ \bigg(\frac{p^{k}~{[n+1-k]}_{p,q}}{{[n+1]}_{p,q}}\bigg)~{\bf{ P_k}}
\end{equation*}

where $k=0,1,2,...,n+1$ and ${\bf{P_{-1}}}={\bf{P_{n+1}}}=0.$\\

 When $ p = 1,$ formula \ref{ee15} reduce to the degree evaluation formula
of the $q$-B$\acute{e}$zier curves. If we let $ P = (P_0, P_1, . . . , P_n)^{T}$  denote the vector of control points of the initial Lupa\c{s} $(p,q)$-B$\acute{e}$zier
curve of degree $n,$ and $ {\bf{P^{(1)}}}=(P_0^\ast, P_1^\ast, . . . , P_{n+1}^\ast)$
 the vector of control points of the degree elevated Lupa\c{s} $(p,q)$-B$\acute{e}$zier curve of
degree $n + 1,$ then we can represent the degree elevation procedure as:

$${\bf{P^{(1)}}}=T_{n+1}{\bf{P}}$$
where

\begin{align*}
 T_{n+1} &= \\
  & \frac{1}{[n+1]_{p,q}}\begin{bmatrix}
\;[n+1]_{p,q}\; & \;0\; & \;\ldots \; & \;0\; & \;0\; \\
[n+1]_{p,q}-p[n]_{p,q} & p[n] _{p,q} & \ldots & 0 & 0 \\
\vdots & \vdots & \ddots & \vdots & \vdots \\
0 & \ldots & [n+1] _{p,q} -p^{n-1}[2] _{p,q} & p^{n-1}[2] _{p,q} & 0 \\
0 & 0 & \ldots & [n+1] _{p,q}-p^n[1] _{p,q} & p^n[1] _{p,q}\\
\;0\; & \;0\; & \;\ldots & \;0\; & \;[n+1] _{p,q} \;
\end{bmatrix}_{(n+2)\times(n+1)} \\
\end{align*}

For any $ l \in \mathbb{N},$ the vector of control points of the degree elevated Lupa\c{s} $(p,q)$-B$\acute{e}$zier curves of degree $ n + l$ is:
${\bf{P^{(l)}}} = T_{n+l}~ T_{n+2}........ T_{n+1} {\bf{P}}.$
As $l \longrightarrow \infty,$ the control polygon $\bf{P^{(l)}}$ converges to a Lupa\c{s} $(p,q)$-B$\acute{e}$zier curve.

\section{Construction of $(p,q)$-analogue of Lupas operators and its limit form}

In this section, we present $(p,q)$-analogue of Lupa\c{s} Bernstein operators and its limit form as follows:\\

The linear operators $L^{n}_{p,q}:$ $C[0,1] \rightarrow C[0,1]$
\begin{equation}\label{e9a}
L^{n}_{p,q}(f;x)= \sum\limits_{k=0}^{n}~~ \frac{f \bigg(\frac{p^{n-k}~[k]_{p,q}}{[n]_{p,q}}\bigg) ~\left[\begin{array}{c}
n \\
k%
\end{array}%
\right] _{p,q} p^{\frac{(n-k)(n-k-1)}{2}} q^{\frac{k(k-1)}{2}}~x^{k}~(1-x)^{n-k}}{\prod\limits_{j=1}^{n}\{p^{j-1}(1-x)+q^{j-1} x\}},
\end{equation}

is $(p,q)$-analogue of Lupa\c{s} Bernstein operators.

Again when $p=1,$  Lupa\c{s} $(p,q)$-Bernstein operators turns out to be Lupa\c{s} $q$-Bernstein operators as given in \cite{mahmudov1,sofia}.\\

It follows directly from the definition that operators $L^{n}_{p,q}(f,t)$ posses the end point interpolation property, that is
$$ L^{n}_{p,q}(f,0)=f(0),~~L^{n}_{p,q}(f,1)=f(1)$$ for all $p>q>0$ and all $n=1,2,.....$\\

Now we show that $(p,q)$-analogue of Lupas operator reproduces linear and constant functions.\\

By some simple computation, we have\\

\textbf{Some auxillary results:}\\

(1)$   L^{n}_{p,q}(1, \frac{u}{u+1})=1$\\

(2) $  L^{n}_{p,q}(t, \frac{u}{u+1})=\frac{u}{u+1} $\\

(3)   $L^{n}_{p,q}(t^2, \frac{u}{u+1})=\frac{u}{u+1} \frac{p^{n-1}}{[n]_{p,q}}+ \frac{qu}{u+1} (\frac{qu}{p+qu}) \frac{[n-1]_{p,q}}{[n]_{p,q}}$\\

or equivalently for $x=\frac{u}{u+1}$\\

(1)$   L^{n}_{p,q}(1, x)=1$\\

(2) $  L^{n}_{p,q}(t, x)=x $\\

(3)   $L^{n}_{p,q}(t^2, x)= \frac{p^{n-1}x}{[n]_{p,q}}+ \frac{q^2~ x^2}{p(1-x)+qx}\frac{[n-1]_{p,q}}{[n]_{p,q}}$\\


\textbf{Proof:} (1) is obvious using $(p,q)$-analogue of Lupas Bernstein basis function.

\begin{equation*}
L^{n}_{p,q}(t;\frac{u}{u+1})= \sum\limits_{k=0}^{n}~~ \frac{\left[\begin{array}{c}
n \\
k%
\end{array}%
\right] _{p,q} p^{\frac{(n-k)(n-k-1)}{2}} q^{\frac{k(k-1)}{2}}~t^{k}~(1-t)^{n-k}}{\prod\limits_{j=1}^{n}\{p^{j-1}(1-t)+q^{j-1} t\}}=1,
\end{equation*}

(2)

\begin{align*}
 L^{n}_{p,q}(t;\frac{u}{u+1}) &=\sum\limits_{k=0}^{n}~~ \frac{ \bigg(\frac{p^{n-k}~[k]_{p,q}}{[n]_{p,q}}\bigg) ~\left[\begin{array}{c}
n \\
k%
\end{array}%
\right] _{p,q} p^{\frac{(n-k)(n-k-1)}{2}} q^{\frac{k(k-1)}{2}}~t^{k}~(1-t)^{n-k}}{\prod\limits_{j=1}^{n}\{p^{j-1}(1-t)+q^{j-1} t\}} \\
   &= \sum\limits_{k=0}^{n}~~ \frac{ \bigg(\frac{p^{n-k}~[k]_{p,q}}{[n]_{p,q}}\bigg) ~\left[\begin{array}{c}
n \\
k%
\end{array}%
\right] _{p,q} p^{\frac{(n-k)(n-k-1)}{2}} q^{\frac{k(k-1)}{2}}~t^{k}~(1-t)^{n-k}}{\prod\limits_{j=0}^{n-1}\{p^{j}(1-t)+q^{j} t\}}\\
&= \sum\limits_{k=0}^{n}~~ \frac{{p^{n-k}} ~\left[\begin{array}{c}
n-1 \\
k-1%
\end{array}%
\right] _{p,q} p^{\frac{(n-k)(n-k-1)}{2}} q^{\frac{k(k-1)}{2}}~t^{k}~(1-t)^{n-k}}{\prod\limits_{j=0}^{n-1}\{p^{j}(1-t)+q^{j} t\}}\\
&= \sum\limits_{k=1}^{n}~~ \frac{{p^{n-k}} ~\left[\begin{array}{c}
n-1 \\
k-1%
\end{array}%
\right] _{p,q} p^{\frac{(n-k)(n-k-1)}{2}} q^{\frac{k(k-1)}{2}}~u^{k}}{\prod\limits_{j=0}^{n-1}\{p^{j}+q^{j} u\}}
\end{align*}

\begin{align*}
 \Rightarrow L^{n}_{p,q}(t;\frac{u}{u+1}) &= \sum\limits_{k=0}^{n-1}~~ \frac{{p^{n-k-1}} ~\left[\begin{array}{c}
n-1 \\
k%
\end{array}%
\right] _{p,q} p^{\frac{(n-k-1)(n-k-2)}{2}} q^{\frac{k(k+1)}{2}}~u^{k+1}}{\prod\limits_{j=0}^{n-1}\{p^{j}+q^{j} u\}}\\
&= \frac{u}{u+1}\sum\limits_{k=0}^{n-1}~~ \frac{{p^{n-1}} ~\left[\begin{array}{c}
n-1 \\
k%
\end{array}%
\right] _{p,q} p^{\frac{(n-k-1)(n-k-2)}{2}} q^{\frac{k(k-1)}{2}}~{(\frac{qu}{p})}^{k}}{\prod\limits_{j=0}^{n-2}\{p^{j} p+q^{j} (qu)\}}\\
&= \frac{u}{u+1}\sum\limits_{k=0}^{n-1}~~ \frac{ ~\left[\begin{array}{c}
n-1 \\
k%
\end{array}%
\right] _{p,q} p^{\frac{(n-k-1)(n-k-2)}{2}} q^{\frac{k(k-1)}{2}}~{(\frac{qu}{p})}^{k}}{\prod\limits_{j=0}^{n-2}\{p^{j} +q^{j} (\frac{qu}{p}u)\}}\\
&= \frac{u}{u+1}
\end{align*}

or equivalently for $x=\frac{u}{u+1}$\\

$L^{n}_{p,q}(t, x)=x.$\\

(3) To prove, $L^{n}_{p,q}(t^2, x)= \frac{p^{n-1}x}{[n]_{p,q}}+ \frac{q^2~ x^2}{p(1-x)+qx}\frac{[n-1]_{p,q}}{[n]_{p,q}}.$\\

Consider

\begin{align*}
 L^{n}_{p,q}(t^2;\frac{u}{u+1}) &=\sum\limits_{k=0}^{n}~~ \frac{ {\bigg(\frac{p^{n-k}~[k]_{p,q}}{[n]_{p,q}}\bigg)}^2 ~\left[\begin{array}{c}
n \\
k%
\end{array}%
\right] _{p,q} p^{\frac{(n-k)(n-k-1)}{2}} q^{\frac{k(k-1)}{2}}~u^{k}}{\prod\limits_{j=0}^{n-1}\{p^{j}+q^{j} u\}} \\
&= \sum\limits_{k=0}^{n}~~ \frac{{p^{2n-2k}}~\frac{[k]_{p,q}}{[n]_{p,q}} ~\left[\begin{array}{c}
n-1 \\
k-1%
\end{array}%
\right] _{p,q} p^{\frac{(n-k)(n-k-1)}{2}} q^{\frac{k(k-1)}{2}}~u^{k}}{\prod\limits_{j=0}^{n-1}\{p^{j}+q^{j} u\}}\\
&= \frac{u}{u+1}\sum\limits_{k=0}^{n-1}~~ p^{n-1-k}~ \frac{[k+1]_{p,q}}{[n]_{p,q}}~ \frac{ ~\left[\begin{array}{c}
n-1 \\
k%
\end{array}%
\right] _{p,q} p^{\frac{(n-k-1)(n-k-2)}{2}} q^{\frac{k(k-1)}{2}}~{(\frac{qu}{p})}^{k}}{\prod\limits_{j=0}^{n-2}\{p^{j} +q^{j} (\frac{q}{p}u)\}}
\end{align*}

\begin{align*}
\Rightarrow L^{n}_{p,q}(t^2;\frac{u}{u+1}) &= \frac{u}{u+1} \frac{[n-1]_{p,q}}{[n]_{p,q}} \sum\limits_{k=0}^{n-1}~~ p^{n-1-k}~ \bigg(\frac{p^k+q[k]_{p,q}}{[n-1]_{p,q}}\bigg)~ \frac{ ~\left[\begin{array}{c}
n-1 \\
k%
\end{array}%
\right] _{p,q} p^{\frac{(n-k-1)(n-k-2)}{2}} q^{\frac{k(k-1)}{2}}~{(\frac{qu}{p})}^{k}}{\prod\limits_{j=0}^{n-2}\{p^{j} +q^{j} (\frac{q}{p}u)\}}\\
&=\frac{u}{u+1}~\frac{p^{n-1}}{[n]_{p,q}}+\frac{qu}{u+1}\frac{[n-1]_{p,q}}{[n]_{p,q}}\sum\limits_{k=0}^{n-1}~~ p^{n-k-1}~ ~ \frac{ ~\left[\begin{array}{c}
n-2 \\
k-1%
\end{array}%
\right] _{p,q} p^{\frac{(n-k-1)(n-k-2)}{2}} q^{\frac{k(k-1)}{2}}~{(\frac{qu}{p})}^{k}}{\prod\limits_{j=0}^{n-2}\{p^{j} +q^{j} (\frac{q}{p}u)\}}\\
&=\frac{u}{u+1}~\frac{p^{n-1}}{[n]_{p,q}}+\frac{qu}{u+1}\frac{[n-1]_{p,q}}{[n]_{p,q}}~\frac{\frac{qu}{p}}{1+\frac{qu}{p}}~\sum\limits_{k=0}^{n-2}~~ p^{n-2}~ ~ \frac{ ~\left[\begin{array}{c}
n-2 \\
k%
\end{array}%
\right] _{p,q} p^{\frac{(n-k-2)(n-k-3)}{2}} q^{\frac{k(k-1)}{2}}~{(\frac{q^2u}{p^2})}^{k}}{p^{n-2}\prod\limits_{j=0}^{n-3}\{p^{j} +q^{j} (\frac{q^2}{p^2}u)\}}\\
&=\frac{u}{u+1}~ \frac{p^{n-1}}{[n]_{p,q}}+ \frac{qu}{u+1} \bigg(\frac{qu}{p+qu}\bigg) \frac{[n-1]_{p,q}}{[n]_{p,q}}
\end{align*}

%

\begin{thm}\label{tp31}
 Let $0<q_n<p_n\leq1$ such that $\lim\limits_{n\to\infty}p_n=1,$
 $\lim\limits_{n\to\infty}q_n=1$ and $\lim\limits_{n\to\infty}p_n^n=a$ $\lim\limits_{n\to\infty}q_n^n=b$ with $0<a,b\leq1$. Then for each $f\in C[0,1],~L^{n}_{p,q}(f;x)$
converges uniformly to $f$ on $C[0,1]$.
\end{thm}

\textbf{Proof:}   Proof is obvious using the following Korovkin's theorem.\\

 \parindent=8mm Let $(T_{n})$ be a sequence of positive linear operators from
$\mathcal{C}[a,b]$ into $\mathcal{C}[a,b].$ Then $\lim_{n}\Vert
T_{n}(f,x)-f(x)\Vert _{\mathcal{C}[a,b]}=0$, for all $f\in \mathcal{C}[a,b]$
if and only if $\lim_{n}\Vert T_{n}(f_{i},x)-f_{i}(x)\Vert _{\mathcal{C}%
[a,b]}=0$, for $i=0,1,2$, where $f_{0}(t)=1,$ $f_{1}(t)=t$ and $%
f_{2}(t^2)=t^{2}.$\\

\parindent=0mm\textbf{Remark:}

\parindent=8mm $~~~~~~~$For $q\in(0,1)$ and $p\in(q,1]$ it is obvious that
$\lim\limits_{n\to\infty}[n]_{p,q}=0 $ or $\frac1{p-q}$. In order to reach to convergence
results of the operator $L^{n}_{p,q}(f;x),$ we take a sequence $q_n\in(0,1)$ and $p_n\in(q_n,1]$
such that $\lim\limits_{n\to\infty}p_n=1,$ $\lim\limits_{n\to\infty}q_n=1$ and $\lim\limits_{n\to\infty}p_n^n=a,$ $\lim\limits_{n\to\infty}q_n^n=b$ with $0<a,b\leq1$. So we get
$\lim\limits_{n\to\infty}[n]_{p_n,q_n}=\infty$.\\

\textbf{$(p,q)$-analogue of the limit Lupa\c{s} Bernstein operators}\\

First, let $0<q<p<1$ and $\lim\limits_{n}  \frac{p^{n-k}~[k]_{p,q}}{[n]_{p,q}}=1-{(\frac{q}{p})}^{k}$

We define linear operators $L^{\infty}_{p,q}$ defined on $C[0,1]$ as

\begin{equation}\label{e9aa}
L^{\infty}_{p,q}(f;x)=\left\{
\begin{array}{ll}
\sum\limits_{k=0}^{\infty}~~ f{(1-{(\frac{q}{p})}^{k}})~b^{k,\infty}_{p,q},~~~~~~~~~~~~~\mbox{if  $x\in[0,1)$ } &  \\
&  \\
f(1),~~~~~~~~~~~~~~~~~~~~~~~~~~~~~~~~~~~~\mbox{if $x=1,$}  &
\end{array}%
\right.
\end{equation}
which is $(p,q)$-analogue of the limit Lupa\c{s} Bernstein operators.

For more details on Limit Lupa\c{s} $q$-analogue of Bernstein operators, one can refer \cite{mahmudov1,sofia}.\\

Note that the function $L^{\infty}_{p,q}(f;x)$ is well-defined on $[0, 1]$ whenever
$f(x)$ is bounded on $[0, 1].$\\

\begin{thm}
 Let $f\in C[0,1],$ $g(x)=f(1-x).$ Then for any $p>q>0,$
\begin{equation*}
L^{n}_{p,q}(f;t)=L^{n}_{\frac{1}{p},\frac{1}{q}}(g;1-t), ~~~~~for~ t\in[0,1]
\end{equation*}
\end{thm}

\textbf{Proof:}

 The proof of above theorem  follow easily along the lines of \cite{sofia} and using the following relations,

 $$[n]_{p,q}={(pq)}^{n-1}[n]_{\frac{1}{q},\frac{1}{p}},$$

$$\frac{p^k [n-k]_{p,q}}{[n]_{p,q}}=1- \frac{p^{k-n}[k]_{\frac{1}{q},\frac{1}{p}}}{[n]_{\frac{1}{q},\frac{1}{p}}}$$ and

$$\left[
\begin{array}{c}
n \\
k%
\end{array}%
\right] _{p,q}=\left[
\begin{array}{c}
n \\
k%
\end{array}%
\right] _{\frac{1}{q},\frac{1}{p}} \frac{{(pq)}^{\frac{k(2n-1-k)}{2}}}{(pq)^{\frac{k(k-1)}{2}}}.$$

\textbf{Remark:} For $p=1,$  this equality coincides with formula (10), Theorem 3 in \cite{sofia}.\\

\textbf{Corollary 1:} Let $p \neq 1$ be fixed, $f\in C[0,1],$ and  $g(x)=f(1-x).$ Then, for $x \in [0,1],$ $L^{n}_{p,q}(f;t)$

converges uniformly to $L^{\infty}_{p,q}(f;t)$  for any $p>0$ being fixed where

\begin{equation}
L^{\infty}_{p,q}(f;t)= \left\{
\begin{array}{ll}L^{\infty}_{p,q}(f;t),~~~~~~~~~~~~~\mbox{if  $0<q<p<1$ } &  \\
&  \\
L^{\infty}_{\frac{1}{q},\frac{1}{p}}(f;t),~~~~~~~~~~~~~\mbox{if $p>q>1$}  &
\end{array}%
\right.
\end{equation}

An explicit form of the limit
function for $0<q<p<1$ is given by \ref{e9aa}. \\

%
%

\begin{thm}
Let $p>q>0,$ $p \neq 1$ be fixed and $f \in C[0,1].$ Then
 $L^{\infty}_{p,q}(f;t)= f(t)$ for all  $t \in [0,1]$ if and only if $f(x)=ax+b$ for some $a,~ b \in \mathbb{R}.$
\end{thm}

\textbf{Corrollary 2:} Operators $L^{\infty}_{p,q}(f;x)$  reproduces linear functions, that is  $L^{\infty}_{p,q}(at+b;x)= ax+b$ for all $p>q>0$  and all $n=1,2,....$ where $a,~ b \in \mathbb{R}.$\\

\textbf{Note:}
$(p,q)$-analogue of Lupas Bernstein operator have an advantage of generating positive linear operators for all $p> q > 0,$ whereas
$(p,q)$-analogue of Bernstein polynomials introduced by Mursaleen et al. generate positive linear operators only if $0<q<p<1.$\\

\section{ $(p,q)$-de Casteljau algorithm for Lupa\c{s} B$\acute{e}$zier curves :}

Lupa\c{s} $(p,q)$-B$\acute{e}$zier curves of degree $n$ can be written as two kinds of linear combination of two Lupa\c{s} $(p,q)$-B$\acute{e}$zier curves of degree $n-1,$ and we can get the two selectable algorithms to evaluate Lupa\c{s} $(p,q)$-B$\acute{e}$zier curves. The algorithms can be expressed as:

\textbf{ Algorithm 1.}\\

\begin{equation}\label{e14}
\left\{
 \begin{array}{ll}
 {\bf{P^{0}_{i}}}(t;p,q)\equiv {\bf{P^{0}_{i}}}\equiv {\bf{P_{i}}}~~~i=0,1,2......,n~~~\mbox{ } &  \\
 &  \\
{\bf{P^{r}_{i}}}(t;p,q)=\frac{q^{n-r}~t}{p^{n-r}(1-t)+q^{n-r} t}~{\bf{P^{r-1}_{i+1}}}(t;p,q)+\frac{p^{n-r}(1-t)}{p^{n-r}(1-t)+q^{n-r} t}~{\bf{P^{r-1}_{i}}}(t;p,q)~~~\mbox{  } &\\
 r=1,...,n,~~~i=0,1,2......,n-r.,~~~\mbox{  }
 \end{array}%
 \right.
 \end{equation}
or

\begin{equation}\label{e15}
\left\{
 \begin{array}{ll}
 {\bf{P^{0}_{i}}}(t;p,q)\equiv {\bf{P^{0}_{i}}}\equiv {\bf{P_{i}}}~~~i=0,1,2......,n~~~\mbox{ } &  \\
 &  \\
{\bf{P^{r}_{i}}}(t;p,q)=\frac{p^{n-i-r}q^i~t}{p^{n-r}(1-t)+q^{n-r} t}~{\bf{P^{r-1}_{i+1}}}(t;p,q)+\frac{p^{n-i-r}~q^i(1-t)}{p^{n-r}(1-t)+q^{n-r} t}~{\bf{P^{r-1}_{i}}}(t;p,q)~~~\mbox{  } &\\
 r=1,...,n,~~~i=0,1,2......,n-r.,~~~\mbox{  }
 \end{array}%
 \right.
 \end{equation}

Then

 \begin{equation}\label{}
  {\bf{ P}}(t; p,q) = \sum\limits_{i=0}^{n-1} {\bf{P_i^1}}(t; p,q)=...=\sum\limits {\bf{P_i^r}}(t; p,q)~ b^{i,{n-r}}_{p,q}(t)=...= {\bf{P_0^n}}~(t; p,q)
\end{equation}

It is clear that the results can be obtained from Theorem (\ref{khalidthm}).  When $p = 1,$ formula (\ref{e14}) and (\ref{e15}) recover the de Casteljau algorithms of classical $q$-B$\acute{e}$zier curves. Let $P^0 = (P_0, P_1, . . . , P_n)^T$ , $P^r = (P_0^r,P_1^r,....,P_{n-r}^r)^{T},$
 then de Casteljau algorithm can be expressed as:\\

\textbf{ Algorithm 2.}

\begin{equation}\label{e16}
 {\bf{ P^r}}(t; p,q)=M_r(t; p,q)....M_2(t; p,q)M_1(t; p,q){\bf{ P^0}}
\end{equation}
where $M_r(t; p,q)$ is a $(n - r + 1) \times (n - r + 2) $ matrix and

$$ M_r(t; p,q)=\begin{bmatrix}
\;\frac{p^{n-r}(1-t)}{p^{n-r}(1-t)+q^{n-r} t}\; & \;\frac{q^{n-r}t}{p^{n-r}(1-t)+q^{n-r} t}\; & \;\ldots \; & \;0\; & \;0\; \\
0 & \frac{p^{n-r}(1-t)}{p^{n-r}(1-t)+q^{n-r} t} & \frac{q^{n-r}t}{p^{n-r}(1-t)+q^{n-r} t} & 0 & 0 \\
\vdots & \vdots & \ddots & \vdots & \vdots \\
0 & \ldots & \frac{p^{n-r}(1-t)}{p^{n-r}(1-t)+q^{n-r} t} &\frac{q^{n-r}t}{p^{n-r}(1-t)+q^{n-r} t} & 0 \\
0 & 0 & \ldots & \frac{p^{n-r}(1-t)}{p^{n-r}(1-t)+q^{n-r} t} & \frac{q^{n-r}t}{p^{n-r}(1-t)+q^{n-r} t}
\end{bmatrix}$$
or
$$ M_r(t; p,q)=\begin{bmatrix}
\;\frac{p^{n-r}(1-t)}{p^{n-r}(1-t)+q^{n-r} t}\; & \;\frac{p^{n-r}t}{p^{n-r}(1-t)+q^{n-r} t}\; & \;\ldots \; & \;0\; & \;0\; \\
0 & \frac{p^{n-r-1}~q(1-t)}{(1-t)+q^{n-r} t} & \frac{p^{n-r-1}~qt}{p^{n-r}(1-t)+q^{n-r} t} & 0 & 0 \\
\vdots & \vdots & \ddots & \vdots & \vdots \\
0 & \ldots & \frac{p~q^{n-r-1}(1-t)}{p^{n-r}(1-t)+q^{n-r} t} &\frac{pq^{n-r-1}t}{p^{n-r}(1-t)+q^{n-r} t} & 0 \\
0 & 0 & \ldots & \frac{q^{n-r}(1-t)}{p^{n-r}(1-t)+q^{n-r} t} & \frac{q^{n-r}t}{p^{n-r}(1-t)+q^{n-r} t}
\end{bmatrix}$$

\section{Tensor product Lupa\c{s} $(p,q)$-B$\acute{e}$zier surfaces on $[0, 1] \times  [0, 1]$}

We define a two-parameter family ${{\bf{P}}}(u,v)$ of tensor product surfaces of degree $m \times n$ as follow:

\begin{equation}\label{e18}
{{\bf{P}}}(u,v) = \sum\limits_{i=0}^{m}\sum\limits_{j=0}^{n} {{\bf{P}}_{i,j}}~ b^{i,m}_{p_1,q_1}(u)~~ b^{j,n}_{p_2,q_2}(v),~~(u,v) \in [0, 1] \times  [0, 1],
\end{equation}

where ${{\bf{P}}_{i,j}} \in \mathbb{R}^3 ~~(i = 0, 1, . . . ,m, j = 0, 1, . . . , n)$ and two real numbers $p_1> q_1>0,~ p_2> q_2 > 0,$ $b^{i,m}_{p_1,q_1}(u),~~ b^{j,n}_{p_2,q_2}(v)$ are Lupa\c{s} $(p,q)$-analogue of Bernstein functions respectively with the parameter $ p_1,q_1$ and $ p_2, q_2.$ We call the parameter surface tensor product
Lupa\c{s} $(p,q)$-B$\acute{e}$zier surface with degree $m \times n.$ We refer to the ${{\bf{P}}_{i,j}}$ as the control points. By joining up adjacent points in the same
row or column to obtain a net which is called the control net of tensor product Lupa\c{s} $(p,q)$-B$\acute{e}$zier surface.\\

\subsection{Properties}

1. \textbf{Geometric invariance and affine invariance property:} Since
\begin{equation}\label{}
 \sum\limits_{i=0}^{m}\sum\limits_{j=0}^{n}  b^{i,m}_{p_1,q_1}(u)~~ b^{j,n}_{p_2,q_2}(v)=1,
\end{equation}

 ${{\bf{P}}}(u,v)$ is an affine combination
of its control points.\\

2. \textbf{Convex hull property:} ${{\bf{P}}}(u,v)$ is a convex combination of ${{\bf{P}}_{i,j}}$ and lies in the convex hull of its control net.\\

3. \textbf{Isoparametric curves property:} The isoparametric curves $v = v^\ast$ and $u = u^\ast$ of a tensor product Lupa\c{s} $(p,q)$-B$\acute{e}$zier surface
are respectively the Lupa\c{s} $(p,q)$-B$\acute{e}$zier curves of degree $m$ and degree $n,$ namely,

\begin{equation*}
{{\bf{P}}}(u,v^\ast) = \sum\limits_{i=0}^{m}\bigg(\sum\limits_{j=0}^{n} {{\bf{P}}_{i,j}}~b^{j,n}_{p_2,q_2}(v^\ast)\bigg)~ b^{i,m}_{p_1,q_1}(u)~~ ,~~u  \in [0, 1] ;
\end{equation*}

\begin{equation*}
{{\bf{P}}}(u^\ast,v) = \sum\limits_{j=0}^{n}\bigg(\sum\limits_{i=0}^{m} {{\bf{P}}_{i,j}}~b^{j,n}_{p_1,q_1}(u^\ast)\bigg)~ b^{i,m}_{p_2,q_2}(v)~~ ,~~v  \in [0, 1]
\end{equation*}

The boundary curves of ${{\bf{P}}}(u,v)$ are evaluated by ${{\bf{P}}}(u,0)$, ${{\bf{P}}}(u,1)$, ${{\bf{P}}}(0,v)$ and ${{\bf{P}}}(1,v)$.\\

4. \textbf{Corner point interpolation property: }The corner control net coincide with the four corners of the surface. Namely, ${{\bf{P}}}(0,0)={{\bf{P}}}_{0,0},$
${{\bf{P}}}(0,1) ={{\bf{P}}}_{0,n},$
 ${{\bf{P}}}(1,0) ={{\bf{P}}}_{m,0},$
 ${{\bf{P}}}(1,1) ={{\bf{P}}}_{m,n},$\\

5. \textbf{Reducibility:} When $p_1 = p_2 = 1,$ formula (\ref{e18}) reduces to a tensor product $q$-B$\acute{e}$zier patch.

\subsection{Degree elevation and $(p,q)$-de Casteljau algorithm for Lupa\c{s} B$\acute{e}$zier surface}

Let ${{\bf{P}}}(u,v)$ be a tensor product Lupa\c{s} $(p,q)$-B$\acute{e}$zier surface of degree $m \times n.$ As an example, let us take obtaining the same
surface as a surface of degree $(m + 1) \times (n + 1).$ Hence we need to find new control points ${{\bf{P}}}_{i,j}^\ast$ such that

\begin{equation}\label{}
{{\bf{P}}}(u,v) = \sum\limits_{i=0}^{m}\sum\limits_{j=0}^{n} {{\bf{P}}_{i,j}}~ b^{i,m}_{p_1,q_1}(u)~~ b^{j,n}_{p_2,q_2}(v)= \sum\limits_{i=0}^{m+1}\sum\limits_{j=0}^{n+1} {\bf{P^\ast}_{i,j}}~ b^{i,m+1}_{p_1,q_1}(u)~~ b^{j,n+1}_{p_2,q_2}(v)
\end{equation}
Let
$\alpha_i=1-\frac{p_1^{i}~{[m+1-i]}_{p_1,q_1}}{{[m+1]}_{p_1,q_1}},~~$ $\beta_j=1-\frac{p_2^{j}~{[n+1-j]}_{p_2,q_2}}{{[n+1]}_{p_2,q_2}}.$

Then
\begin{equation}\label{}
  {{\bf{P}}}_{i,j}^\ast=\alpha_i~\beta_j~{{\bf{P}}}_{i-1,j-1}+\alpha_i~(1-\beta_j)~{{\bf{P}}}_{i-1,j}+(1-\alpha_i)~\beta_j~{{\bf{P}}}_{i,j-1}+(1-\alpha_i)~(1-\beta_j)~{{\bf{P}}}_{i,j}
\end{equation}

which can be written in matrix form as
$$
  \begin{bmatrix}
    1-\frac{p_1^{i}~{[m+1-i]}_{p_1,q_1}}{{[m+1]}_{p_1,q_1}} & \frac{p_1^{i}~{[m+1-i]}_{p_1,q_1}}{{[m+1]}_{p_1,q_1}}\\
  \end{bmatrix}
    \begin{bmatrix}
      {{\bf{P}}}_{i-1,j-1} & {{\bf{P}}}_{i-1,j} \\
     {{\bf{P}}}_{i,j-1} & {{\bf{P}}}_{i,j} \\
    \end{bmatrix}
                   \begin{bmatrix}
                     1-\frac{p_2^{j}~{[n+1-j]}_{p_2,q_2}}{{[n+1]}_{p_2,q_2}} \\
                     \frac{p_2^{j}~{[n+1-j]}_{p_2,q_2}}{{[n+1]}_{p_2,q_2}}\\
                   \end{bmatrix}
    $$

The de Casteljau algorithms are also easily extended to evaluate points on a Lupa\c{s} $(p,q)$-B$\acute{e}$zier surface. Given the control net
${{\bf{P}}}_{i,j} \in \mathbb{R}^3, i = 0, 1, . . . ,m,~~ j = 0, 1, . . . , n.$

\begin{equation}\label{e21}
\left\{
 \begin{array}{ll}
 {\bf{P^{0,0}_{i,j}}}(u,v)\equiv {\bf{P^{0,0}_{i,j}}}\equiv {\bf{P_{i,j}}}~~~i=0,1,2......,m;~~j=0,1,2...n.\mbox{ } &  \\
 &  \\
{\bf{P^{r,r}_{i,j}}}(u,v)= \begin{bmatrix}
    \frac{p_1^{m-r}(1-u)}{p_1^{m-r}(1-u)+q_1^{m-r} u} & \frac{q_1^{m-r}~u}{p_1^{m-r}(1-u)+q_1^{m-r}u}\\
  \end{bmatrix}
    \begin{bmatrix}
      {{\bf{P}}}_{i,j}^{r-1,r-1} & {{\bf{P}}}_{i,j+1}^{r-1,r-1} \\
     {{\bf{P}}}_{i+1,j}^{r-1,r-1} & {{\bf{P}}}_{i+1,j+1}^{r-1,r-1} \\
    \end{bmatrix}
                   \begin{bmatrix}
                      \frac{p_2^{n-r}(1-v)}{p_2^{m-r}(1-v)+q_2^{n-r} v} \\
                     \frac{q_2^{n-r}~v}{p_2^{n-r}(1-v)+q_2^{n-r}v}\\
                   \end{bmatrix}      ~~~\mbox{  } &\\
 r=1,...,k, ~k=\text{min}(m,n)~~~i=0,1,2......,m-r;~~ j=0,1,....n-r~~~\mbox{  }
 \end{array}%
 \right.
 \end{equation}
 or

\begin{equation}\label{e22}
\left\{
 \begin{array}{ll}
 {\bf{P^{0,0}_{i,j}}}(u,v)\equiv {\bf{P^{0,0}_{i,j}}}\equiv {\bf{P_{i,j}}}~~~i=0,1,2......,m;~~j=0,1,2...n.\mbox{ } &  \\
 &  \\
{\bf{P^{r,r}_{i,j}}}(u,v)= \begin{bmatrix}
    \frac{p_1^{m-i-r}q_1^i(1-u)}{p_1^{m-r}(1-u)+q_1^{m-r} u} & \frac{p_1^{m-i-r}q_1^i~u}{p_1^{m-r}(1-u)+q_1^{m-r}u}\\
  \end{bmatrix}
    \begin{bmatrix}
      {{\bf{P}}}_{i,j}^{r-1,r-1} & {{\bf{P}}}_{i,j+1}^{r-1,r-1} \\
     {{\bf{P}}}_{i+1,j}^{r-1,r-1} & {{\bf{P}}}_{i+1,j+1}^{r-1,r-1} \\
    \end{bmatrix}
                   \begin{bmatrix}
                      \frac{p_2^{n-j-r}q_2^j(1-v)}{p_2^{n-r}(1-v)+q_2^{n-r} v} \\
                     \frac{p_2^{n-j-r}q_2^j~v}{p_v^{n-r}(1-v)+q_2^{n-r}v}\\
                   \end{bmatrix}      ~~~\mbox{  } &\\
 r=1,...,k, ~k=\text{min}(m,n)~~~i=0,1,2......,m-r;~~ j=0,1,....n-r~~~\mbox{  }
 \end{array}%
 \right.
 \end{equation}
 When $m = n,$ one can directly use the algorithms above to get a point on the surface. When $m \neq n,$ to get a point on the
surface after $k$ applications of formula (\ref{e21}) or (\ref{e22}), we perform formula (\ref{e16}) for the intermediate point $ {{\bf{P}}}_{i,j}^{k,k}.$\\

\textbf{Note:} We get Lupa\c{s} $q$-B$\acute{e}$zier curves and surfaces for $(u, v) \in [0, 1] \times [0, 1] $  when we set the parameter $p_1=p_2=1$ as proved in \cite{wcq}.\\


\newpage
\section{Shape control of $(p,q)$-B$\acute{e}$zier curves and surfaces}

We have constructed Lupa\c{s} type $(p,q)$-Bernstein functions which holds the end point interpolation property as shown in following figures of curves and surfaces. It can be also observed that the curves (surfaces) generated is contained within the convex hull of the control polygon (control net) for different values of $ p$ and $q$.\\

Parameter $p$ and $q$ has been used to control the shape of curves and surfaces: if $ 0 < q < p \leq 1,$ as $p$ and $q$ decreases, the curves (surfaces) move close to the control polygon (control net), as $p$ and $q$ increases, the curves (surfaces) move away from the control polygon (control net).\\

If $1<q<p,$ the effects of $p$ and $q$ are opposite, as $p$ and $q$ decreases, the curves (surfaces) move away from the control polygon (control net), as $p$ and $q$ increases, the curves (surfaces) move close to the control polygon (control net) which can be seen in the following figures.

\begin{figure}[htb!]
\begin{subfigure}{.5\textwidth}
\includegraphics[width=1\linewidth, height=6cm]{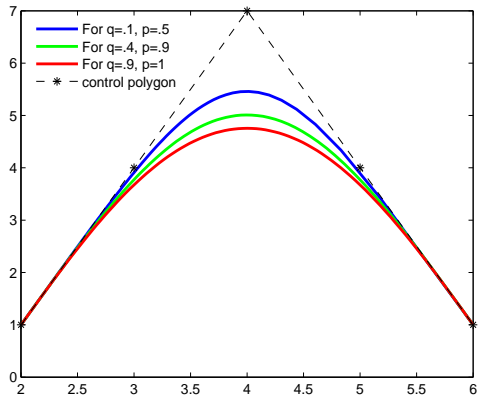}
\caption{$0<q<p\leq1$}\label{f6}
\label{fig:subim1}
\end{subfigure}
\begin{subfigure}{.5\textwidth}
\includegraphics[width=1\linewidth, height=6cm]{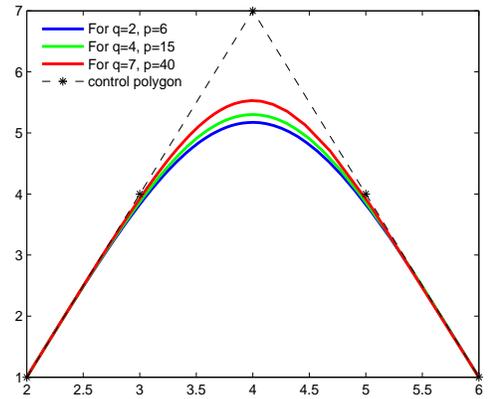}
\caption{$1<q<p$}\label{f7}
\label{fig:subim2}
\end{subfigure}
\begin{subfigure}{.5\textwidth}
\includegraphics[width=1\linewidth, height=6cm]{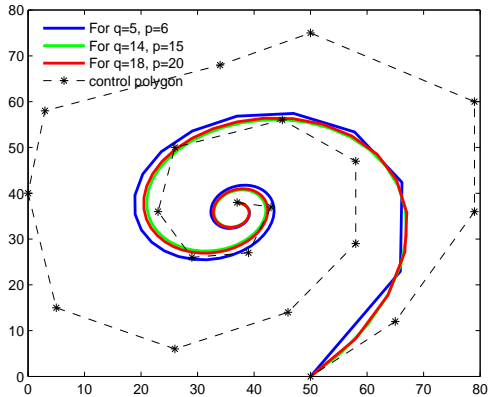}
\caption{$1<q<p$}\label{f9}
\label{fig:subim3}
\end{subfigure}
\caption{The effect of the shape of Lupa\c{s} $(p,q)$-B$\acute{e}$zier curves}
\label{fig:image5}
\end{figure}

\begin{figure}[htb!]
\begin{subfigure}{.5\textwidth}
\includegraphics[width=1\linewidth, height=6cm]{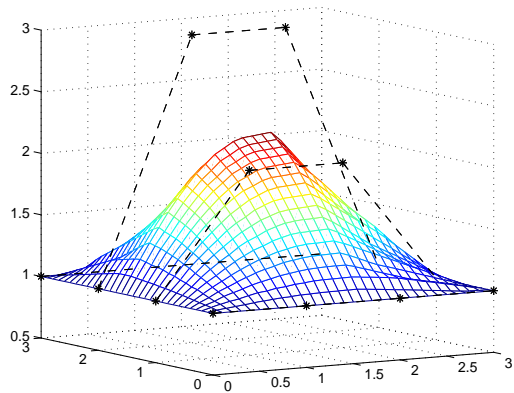}
\caption{$q_1=q_2=.1,$ $p_1=p_2=.7$ }\label{f11}
\label{fig:subim1}
\end{subfigure}
\begin{subfigure}{.5\textwidth}
\includegraphics[width=1\linewidth, height=6cm]{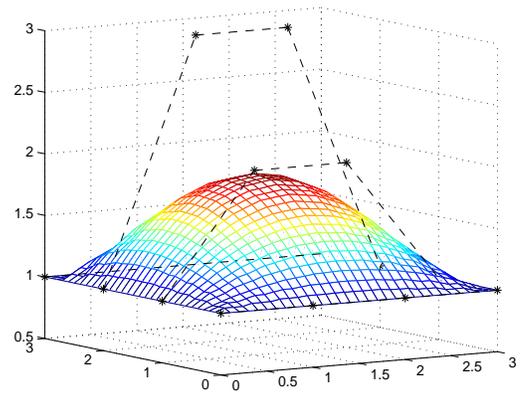}
\caption{$q_1=q_2=.99,$ $p_1=p_2=1$}\label{f10}
\label{fig:subim2}
\end{subfigure}
\begin{subfigure}{.5\textwidth}
\includegraphics[width=1\linewidth, height=6cm]{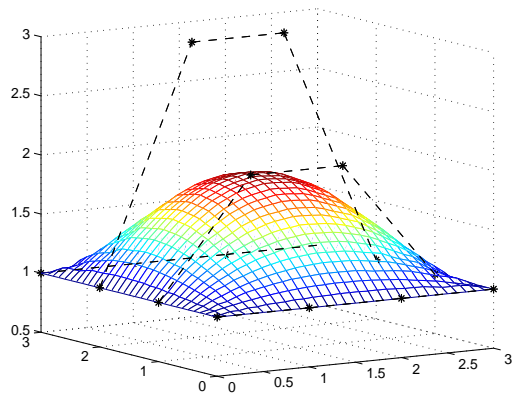}
\caption{$q_1=q_2=3,$ $p_1=p_2=5$}\label{f12}
\label{fig:subim3}
\end{subfigure}
\begin{subfigure}{.5\textwidth}
\includegraphics[width=1\linewidth, height=6cm]{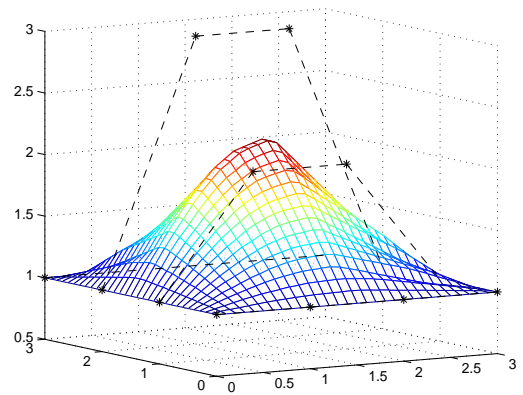}
\caption{$q_1=q_2=20,$ $p_1=p_2=200$}\label{f13}
\label{fig:subim4}
\end{subfigure}
\caption{The effect of the shape of Lupa\c{s} $(p,q)$-B$\acute{e}$zier surfaces}
\label{fig:image5}
\end{figure}

\newpage
\textbf{Acknowledgement:}
We are very grateful to the anonymous referees for the inspiring comments and the valuable suggestions which improved
our paper considerably.


\end{document}